\documentclass[sigconf,authorversion=true]{acmart}

\usepackage[english]{babel}
\usepackage[utf8]{inputenc}
\usepackage[inline]{enumitem}
\usepackage{booktabs}
\usepackage{xcolor}
\usepackage{tikz}
\usepackage{listings}
\usepackage{graphicx}

\definecolor{javared}{rgb}{0.6,0,0}
\definecolor{javagreen}{rgb}{0.25,0.5,0.35}
\definecolor{javapurple}{rgb}{0.5,0,0.35}
\definecolor{javadocblue}{rgb}{0.25,0.35,0.75}
\newcommand{\jkCodeEntry}[1]{\texttt{#1}}

\begin{document}

\copyrightyear{2018}
\acmYear{2018}
\setcopyright{acmlicensed}
\acmConference[ManLang'18]{15th International Conference on Managed Languages & Runtimes}{September 12--14, 2018}{Linz, Austria}
\acmBooktitle{15th International Conference on Managed Languages \& Runtimes (ManLang'18), September 12--14, 2018, Linz, Austria}
\acmPrice{15.00}
\acmDOI{10.1145/3237009.3237017}
\acmISBN{978-1-4503-6424-9/18/09}

\title{Debugging Native Extensions of Dynamic Languages}
\titlenote{This research project is partially funded by Oracle Labs.}

\author{Jacob Kreindl}
\affiliation{
  	\institution{Johannes Kepler University Linz}
  	\country{Austria}
}
\email{jacob.kreindl@jku.at}

\author{Manuel Rigger}
\affiliation{
	\institution{Johannes Kepler University Linz}
	\country{Austria}
}
\email{manuel.rigger@jku.at}

\author{Hanspeter M{\"{o}}ssenb{\"{o}}ck}
\affiliation{
	\institution{Johannes Kepler University Linz}
	\country{Austria}
}
\email{hanspeter.moessenboeck@jku.at}

\begin{abstract}
	Many dynamic programming languages such as Ruby and Python enable developers to use so called native extensions, code implemented in typically statically compiled languages like C and C++.
	However, debuggers for these dynamic languages usually lack support for also debugging these native extensions.
	GraalVM can execute programs implemented in various dynamic programming languages and, by using the LLVM-IR interpreter Sulong, also their native extensions.
	We added support for source-level debugging to Sulong based on GraalVM's debugging framework by associating run-time debug information from the LLVM-IR level to the original program code.
	As a result, developers can now use GraalVM to debug source code written in multiple LLVM-based programming languages as well as programs implemented in various dynamic languages that invoke it in a common debugger front-end.
\end{abstract}



\maketitle

\renewcommand{\shortauthors}{J. Kreindl et al.}

\section{Introduction}
\label{sec:introduction}

Tooling support for dynamic languages such as Python, Ruby, and R typically includes one or several debuggers to enhance the developers' experience.
However, applications written in these languages often also invoke \emph{native extensions}, that is, code written in low-level languages such as C/C++ or Fortran.
Existing debuggers for dynamic languages generally lack support for debugging native extensions, forcing programmers to fall back to other debugging approaches.

Existing cross-language debuggers~\cite{rw:blink,rw:python:2,rw:general-1} are either limited to a very specific combination of programming languages or require language implementers to modify preexisting debuggers.
As an alternative, many interpreters for popular high-level languages provide some degree of integration with low-level debuggers like \textit{gdb}~\cite{rw:python:1,rw:r}.
However, these efforts mostly fall short of an integrated debugging experience~\cite{rw:general-2}.
The alternative solution of attaching separate debuggers for high-level and low-level code to the same process requires developers to frequently switch between different front-ends with differing usage concepts.

\sloppy The \emph{GraalVM}\footnote{GraalVM releases are available at \url{https://www.graalvm.org/}.} is an extended \emph{Java Virtual Machine} (JVM) that can execute various programming languages~\cite{graalvm:onevm}.
While the project focuses on dynamic languages such as Ruby, Python, R, and JavaScript, GraalVM's integrated LLVM-IR interpreter, called Sulong~\cite{sulong:asplos}, supports executing LLVM-based languages like C and C++.
By using Sulong, language implementers have been able to efficiently implement native function interfaces.
However, they have not been able to debug native extensions, as Sulong lacked support for GraalVM's integrated debugging framework.
As part of the work described in this paper, we implemented source-level debugging support for LLVM-based languages in Sulong.

LLVM-IR~\cite{llvm} is an intermediate representation of source-code that can be produced by various LLVM front-ends.\footnote{\textit{Clang} (\url{https://clang.llvm.org/}) can compile code in various members of the \textit{C} family of programming languages to LLVM-IR. \textit{DragonEgg} (\url{https://dragonegg.llvm.org/}) can do the same for C/C++/Fortran code.}
LLVM-IR programs can contain \emph{debug information} which relates them to the original program code.
We enriched Sulong's run-time program representation with this data and apply it to reconstruct the state of the original program from the executed LLVM-IR for GraalVM's built-in framework for cross-language, source-level debugging~\cite{instrumentation}.

Sulong improves upon existing approaches for debugging native extensions of dynamic languages.
It enables users to debug their entire program in a single user interface instead of frequently switching between different debuggers and their corresponding usage schemes.
GraalVM's implementation of language interoperability also allows the native debugger to automatically display complex values received from other languages without requiring the user to specify its source language first.
Furthermore, new language implementations in GraalVM that execute native extensions with Sulong gain debugging support for these native extensions without additional programming efforts.

\section{Background}
\label{sec:bg:main}

\begin{figure}
	\centering
	\includegraphics[width=.95\columnwidth,page=1,trim={6.35mm 6.35mm 6.35mm 6.35mm},clip]{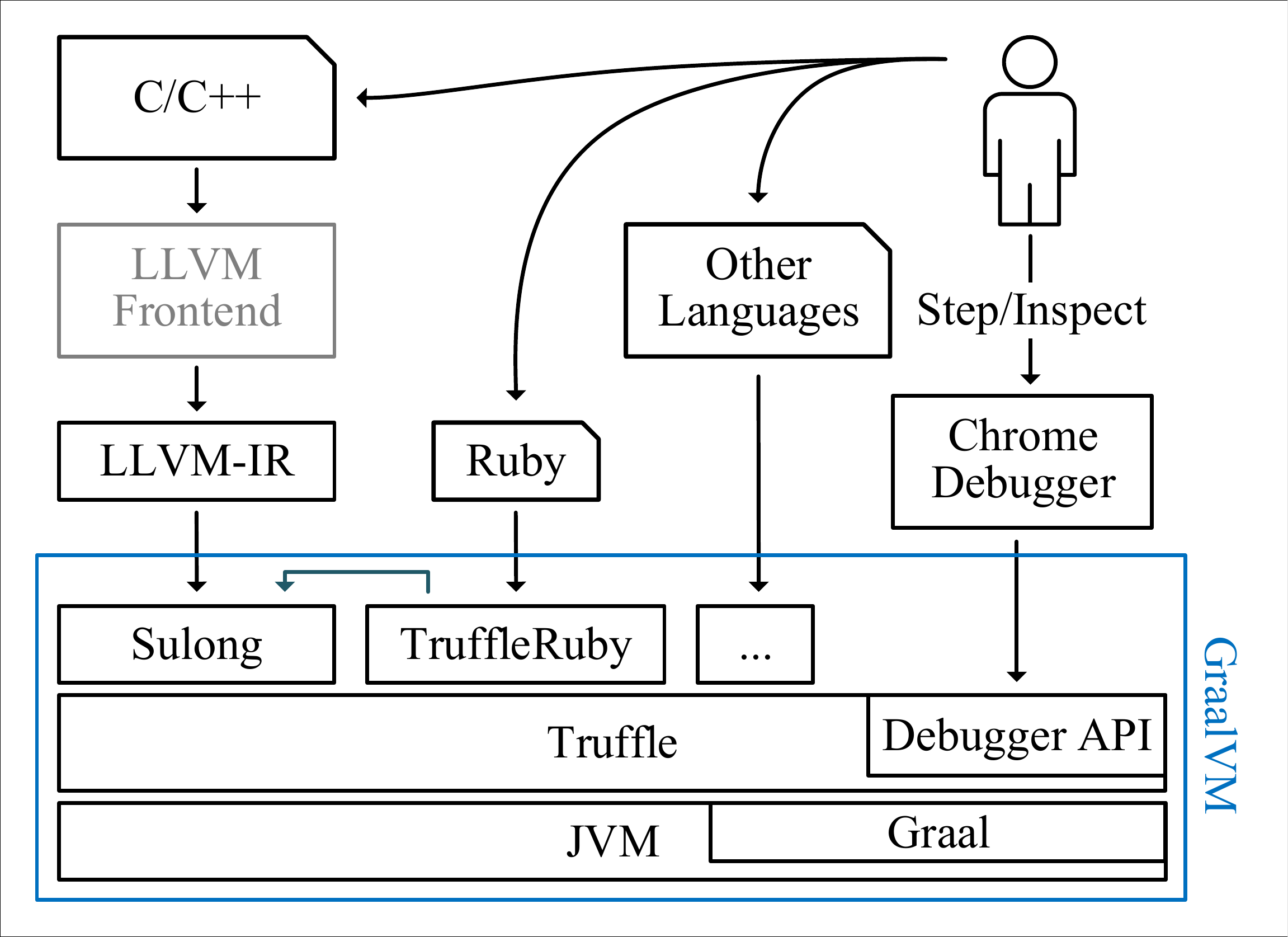}
	\caption{Debugging in GraalVM}
	\label{fig:graalvm}
\end{figure}

Figure~\ref{fig:graalvm} displays the overall structure of our approach.
Developers often implement programs in multiple programming languages.
Many dynamic languages enable this by allowing programmers to invoke \emph{native extensions}, that is, code written in languages such as C and C++ that are typically compiled statically.
\emph{Truffle}~\cite{graalvm:onevm} is a framework for implementing interpreters for programming languages.
Sulong~\cite{sulong:asplos} and TruffleRuby~\cite{truffleruby} are existing Truffle language implementations.
Truffle language implementations can interact by allowing programs to share both functions and values.
TruffleRuby and other Truffle-based interpreters, many of which are also part of GraalVM, use this feature to execute native extensions with Sulong.
Truffle also contains a language-independent framework for source-level debugging~\cite{instrumentation}.
In combination with various front-ends such as the \emph{Chrome Debugger}\footnote{The \textit{Chrome Debugger} is part of the \emph{Chrome DevTools} available at \url{https://developers.google.com/web/tools/chrome-devtools/}.}, it enables developers to debug all of their code in the same user interface.
Below, we explain the components of Figure~\ref{fig:graalvm} in more detail.

\paragraph{Truffle}

Truffle~\cite{graalvm:onevm} is a framework for implementing interpreters for executing programming languages on top of the JVM\@.
Truffle language implementations parse source code into an \emph{Abstract Syntax Tree} (AST) representation, which Truffle can execute.
This language-independent program representation allows different Truffle-based interpreters to share both functions and values.
It also enables the framework to provide a common tooling infrastructure for all Truffle-based language implementations.
This includes a framework for cross-language source-level debugging, which tools such as GraalVM's built-in backend for the Chrome debugger can access using the \emph{Truffle Debugger API}.
Truffle can be used together with the \emph{Graal} dynamic compiler~\cite{graal:basic, graal:ir} to enhance the execution performance of guest-language programs~\cite{graal:fortruffle} and to minimize the run-time overhead of the tooling support~\cite{graalvm:debugging}.
GraalVM includes a JVM with Graal and Truffle as well as multiple Truffle-based languages and tools.

\paragraph{LLVM-IR}

LLVM is a framework for program compilation and optimization~\cite{llvm}.
It provides an intermediate representation of source code called \emph{LLVM-IR}.
We refer to the binary encoding of LLVM-IR as \emph{LLVM bitcode}.
Existing LLVM-IR front-ends such as \emph{Clang}\footnote{\emph{Clang} is available at \url{https://clang.llvm.org/}.}, which can parse code written in various members of the C-family of programming languages, can compile input programs to LLVM bitcode files.
While compiling programs to LLVM-IR, these front-ends can also generate \emph{debug information} and include it in the bitcode files.
This debug information enables debuggers and other tools to relate instructions and symbols in LLVM-IR to the expressions and symbols in the original source code they represent.
LLVM-IR is in \emph{Static Single Assignment} (SSA) form~\cite{ssa} and uses a syntax and instruction set similar to RISC-assembly~\cite{llvm}.

\paragraph{Sulong}

Sulong~\cite{sulong:asplos} is a Truffle-based interpreter for LLVM bitcode programs.
It uses an approach based on dynamic dispatch of basic blocks to support LLVM-IR's unstructured control flow~\cite{sulong:vmil}.
Other programming language interpreters that are also part of GraalVM, e.g.,~\emph{TruffleRuby} for Ruby code and \emph{GraalPython}~\cite{graalpython} for Python code, support executing native extensions compiled to LLVM-IR on Sulong.
Many of these languages also support the Truffle Debugger API.
As part of this paper, we describe how we implemented support for it in Sulong.

\section{Run-Time Debug Information}
\label{sec:debuginformation}

Debug information in LLVM-IR programs relates LLVM-IR instructions to locations in the source code and provides an association of source-level symbols to run-time values.
Sulong attaches an in-memory representation of this debug information to the Truffle AST and its global scope to provide on-demand access to the program's source-level state to Truffle's debugging framework.

\newcommand{\jkStepLabel}[1]{\textcolor{red}{\textcircled{#1}}}

\begin{lstlisting}[float,escapeinside={/*}{*/},language=c,caption={Annotated C-code for the factorial function.},label={lst:fact}]
/*\jkStepLabel{1}*/int fact(int n) {
int result = 1;
if (/*\jkStepLabel{3}*/n /*\jkStepLabel{2}*/> 0)
  /*\jkStepLabel{7}*/result = n/*\jkStepLabel{6}*/*/*\jkStepLabel{5}*/fact(n/*\jkStepLabel{4}*/- 1);
/*\jkStepLabel{8}*/return result;
}
\end{lstlisting}

\begin{figure}
	\centering
	\includegraphics[width=\columnwidth,page=2,trim={6.35mm 6.35mm 6.2mm 6.35mm},clip]{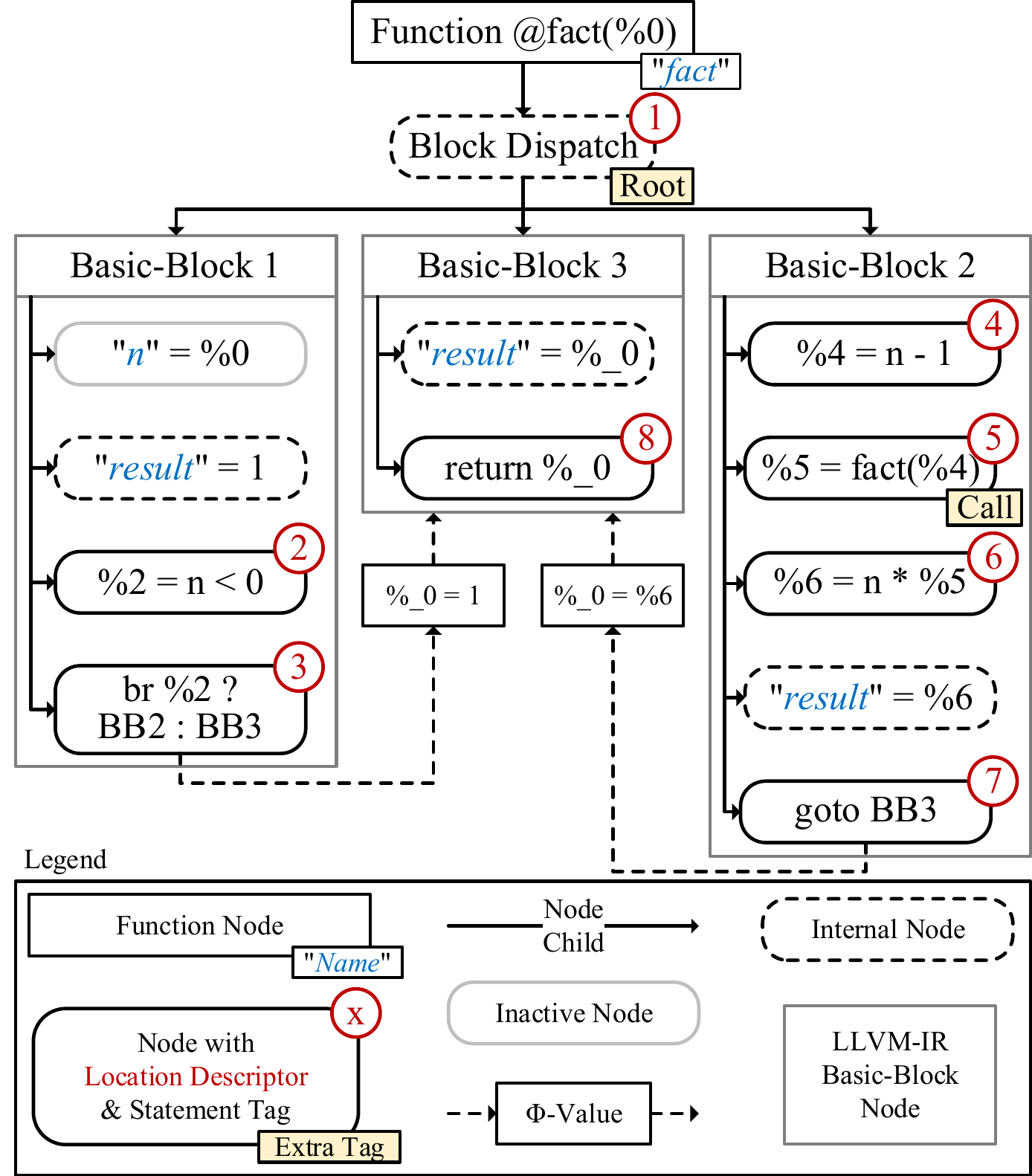}
	\caption{Sulong AST of the factorial function in Listing~\ref{lst:fact}. The program was compiled with Clang 5 and optimized only with \textit{mem2reg}.}
	\label{fig:fact_AST}
\end{figure}

In the following, we will use our implementation of the factorial function, which is shown in Listing~\ref{lst:fact}, to demonstrate how Sulong represents debug information at run-time.
Figure~\ref{fig:fact_AST} illustrates Sulong's Truffle AST for the LLVM-IR produced from the \jkCodeEntry{fact} function in Listing~\ref{lst:fact}.
It shows three \emph{Basic-Block} nodes as children of a \emph{Block Dispatch node}.
The Block Dispatch node transfers control between the individual Basic-Block nodes as directed by the last instruction in each Basic-Block node and sets the value of \jkCodeEntry{\%\_0}, a so-called $\Phi$-Instruction~\cite{llvm,ssa} whose value is determined based on control flow.
For example, in Basic-Block 1, the \jkCodeEntry{br} instruction selects either Basic-Block 2 or Basic-Block 3 as a successor, depending on the boolean value \jkCodeEntry{\%2}.

\subsection{Stepping \& Breakpoints}
\label{sec:di:controlflow}

Truffle's Debugging Framework relies on \emph{tags} and \emph{location descriptors} attached to AST nodes to determine their source-level semantics.
It uses this information present in the AST to support stepping through the original source code and setting breakpoints in it.

\newcommand{\jkSubParagraph}[1]{\noindent{}\textbf{#1}}

\paragraph{Tags}

Truffle's debugging framework requires Truffle language implementations to annotate the ASTs they produce with special \emph{tags}.
These tags enable the debugging framework to implement source-level single-stepping and breakpoints, build source-level call stacks and to unwind them on user request.

\jkSubParagraph{Statements.} The debugging framework defines the \emph{Statement} tag to identify nodes at which it may suspend the executing program to single-step through the original source code.
Sulong attaches this tag to all nodes that represent LLVM-IR instructions for which debug information defines a source-location.
This tagging strategy enables the debugger to step on expressions in the original source code rather than just statements.
In Figure~\ref{fig:fact_AST}, for example, \jkCodeEntry{\%2 = n < 0} is a statement that has the location descriptor \jkStepLabel{2} attached to it.
Users can set breakpoints in the source code to navigate through the program more coarsely than stepping on expression-level.

\jkSubParagraph{Function Roots.} The debugging framework also defines the \emph{Root} tag to mark the entry point to a function's body.
Nodes with this tag act as boundaries for stepping \textit{into} and \textit{out of} function calls.
They also represent locations at which the debugging framework may resume execution of the guest-language program to restart an already executing guest-language function.
As Figure~\ref{fig:fact_AST} shows, Sulong marks the Block Dispatch node with the \emph{Root} tag.
In contrast to the actual AST root, the Block Dispatch node has access to the original values passed as arguments to the function when it was called.
This access enables it to restart the function with the original arguments, though it is incapable of undoing any modifications the function already applied to the native heap.
Since a user application may require considerable time to reach the point in its execution at which the user actually wants to debug it, the ability to re-execute just the function of interest removes an obstacle for users to inspect the function's execution again under a different point of view.

\jkSubParagraph{Function Calls.} Lastly, the debugging framework uses the \emph{Call} tag to identify those AST nodes among the currently executing ones, whose source-level locations the front-end should display in the call-stack alongside the source-level names of the functions they are part of.
Sulong marks all nodes with this tag which represent calls to a source-level function.
Figure~\ref{fig:fact_AST} shows that the node performing the recursive call to \jkCodeEntry{fact} in Basic-Block 2 is marked with the \emph{Call} tag.
\textit{Function} nodes in the AST also provide the source-level name of their corresponding functions to be displayed in the call-stack.
These names are part of debug information and, unlike names of LLVM-IR functions, not mangled for linking.

\paragraph{Location Descriptors}

The debugging framework needs to be able to associate a tagged node with the part of the original program's source code which the node represents.
Sulong attaches this information to statement nodes in the form of \emph{location descriptors}.
Instructions in LLVM-IR usually correspond to distinct expressions in the original source code.
For each of these instructions, debug information in LLVM bitcode files describes the location of the corresponding expressions with an absolute file path as well as a line and column number.
Truffle defines data structures to describe lexical regions within a text source and expects language implementations to provide one such source section for every tagged node.
However, these source sections can only be created for valid locations within accessible files.
If, for example, a user attempts to debug a bitcode program without recompiling any source code they modified, debug information in the bitcode program can reference invalid source location.
We implemented location descriptors for Sulong which reference Truffle's source descriptors but also allow the interpreter to retain location information even for inaccessible sources or invalid locations within accessible sources.
While stepping through the program is not possible in the latter case, Sulong can still use the information provided by these location descriptors to provide stack-traces on errors during guest-language execution.
As Figure~\ref{fig:fact_AST} shows, Sulong attaches one such location descriptor to each node representing a source-level expression.

\subsection{Symbol Inspection}
\label{sec:di:symbolinspection}

Truffle-based debuggers display the current values for all source-level symbols that are defined at the point in the program at which it was suspended by the debugger. Users can inspect these symbols and their values to determine the program's state.

\begin{figure}
	\centering
	\includegraphics[width=.8\columnwidth,page=4,trim={6.35mm 6.35mm 6.35mm 6.35mm},clip]{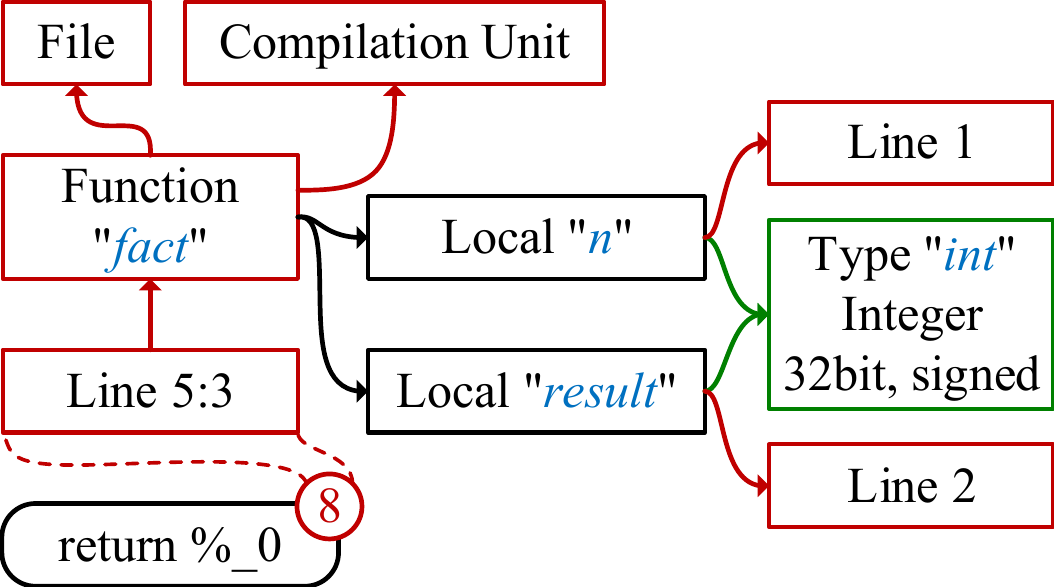}
	\caption{Sulong scope hierarchy at \jkStepLabel{8} in Listing~\ref{lst:fact}.}
	\label{fig:scope}
\end{figure}

\begin{lstlisting}[float,language=llvm,caption={Partial LLVM-IR describing \jkStepLabel{8} in Listing~\ref{lst:fact}.},label={lst:ir-8}]
define i32 @fact(i32) !dbg !7 {
  ; <...>
  ret i32 %.0, !dbg !23
}
!1 = File(name: "fact.c", path: "<...>")
!7 = Subprogram(name: "fact", scope: !1, file: !1, line: 1, <...>)
!10 = BasicType(name: "int", size: 32, encoding: signed_integer)
!11 = LocalVariable(name: "n", scope: !7, line: 1, type: !10, <...>)
!14 = LocalVariable(name: "result", scope: !7, line: 2, type: !10, <...>)
!23 = Location(line: 5, col: 3, scope: !7)

\end{lstlisting}

We implemented descriptors for source-level symbols, scopes and types to represent the corresponding debug information in Sulong.
The interpreter uses this information at run-time to derive a representation of the source-level program state, that is, the values of all local and global symbols in the source code, which it then provides to the debugger framework.
Figure~\ref{fig:scope} illustrates the composition of these descriptors at the \jkCodeEntry{return} statement in Listing~\ref{lst:fact}.
The location descriptor attached to the node references another descriptor representing the \jkCodeEntry{fact} function as its parent scope.
The function scope references the file it was declared in as its parent scope, its source-level name and two symbol descriptors which describe the argument \jkCodeEntry{n} and the local variable \jkCodeEntry{result}.
Both symbols descriptors reference the same type descriptor for C's \jkCodeEntry{int} type and a location that describes their declaration site.
The function also references its compilation unit which, in this case, does not contain any global symbols.
Listing~\ref{lst:ir-8} shows parts of the LLVM-IR and debug information from which Sulong parsed this representation.

\paragraph{Symbols}

The \emph{symbol descriptors} we implemented in Sulong encode all information about source-level named symbols that is required to display their values at run-time.
As shown in Figure~\ref{fig:scope}, these descriptors reference the symbol's name, type and---by a location descriptor---also its declaration site.
They distinguish between dynamic symbols, which are defined and accessible only after source locations lexically succeeding their definition, and static ones, which exist at any point in a function's execution.
Sulong does not provide values for dynamic symbols to the debugger at expression locations preceding their definition.

\paragraph{Types}

The \emph{type descriptors} we implemented for Sulong contain all information required to derive a representation which corresponds to the described source-level type from a run-time value.
We implemented various versions of these type descriptors, each specialized to a different kind of type such as structure, array, pointer, enumeration or primitive.
Each specialized descriptor only stores information necessary to format values of its type.
While an array type references only a single element type and integer length, a structured type such as a C++ class stores a name, type and offset for each of its members, including those declared in any of its parent types.
A primitive type references a binary encoding, while an enumeration type stores a mapping from IR-level values to the labels they correspond to.
Common information among all type descriptors includes the type's name to be displayed by the debugger and its bit-size.

\paragraph{Scopes}

The location descriptors we implemented for Sulong also describe source-level scopes and their hierarchy.
For this reason, they can reference a parent scope as well as an arbitrary number of symbol descriptors.
To keep the memory overhead of these scope descriptors minimal, we defined an abstract interface for them and created various subclasses for this interface, each specialized for a certain kind of scope.
Similar to the \textit{expression} descriptors we discussed in Section~\ref{sec:di:controlflow}, a \textit{symbol} scope describes the declaration site of a named symbol or type member.
In contrast to other scopes, \textit{symbol} and \textit{expression} scopes cannot contain members as they describe a declaration site rather than a semantic scope.
\textit{Blocks} and \textit{functions} describe their lexical entry point in order to enable Sulong to determine whether a function-local symbol in the scope hierarchy is actually defined at an expression at which the debugger suspended the program.
Functions additionally provide their source-level name and reference the \textit{compilation unit} they are contained in to give the user access to the global symbols in that scope.
Sulong attaches such a function descriptor to any AST root node that represents a source-level function.
A \textit{type}, e.g. a C++ class or union, can be the parent scope of source-level instance functions but, like a type descriptor, may also contain symbols which represent static members.
Descriptors for \textit{named scopes}, such as C++ namespaces, store their name, but no lexical region as they may span multiple source files.
If such a named scope is referenced in multiple bitcode files, Sulong uses the same descriptor for each occurence.
This enables Sulong to collect all symbols declared within the scope in the same descriptor, regardless of which compilation unit the symbol declaration was part of.

\section{Source-Level Value Inspection}
\label{sec:valueinspection}

\begin{figure}
	\centering
	\includegraphics[width=.9\columnwidth,page=3,trim={6.35mm 6.35mm 6.35mm 6.35mm},clip]{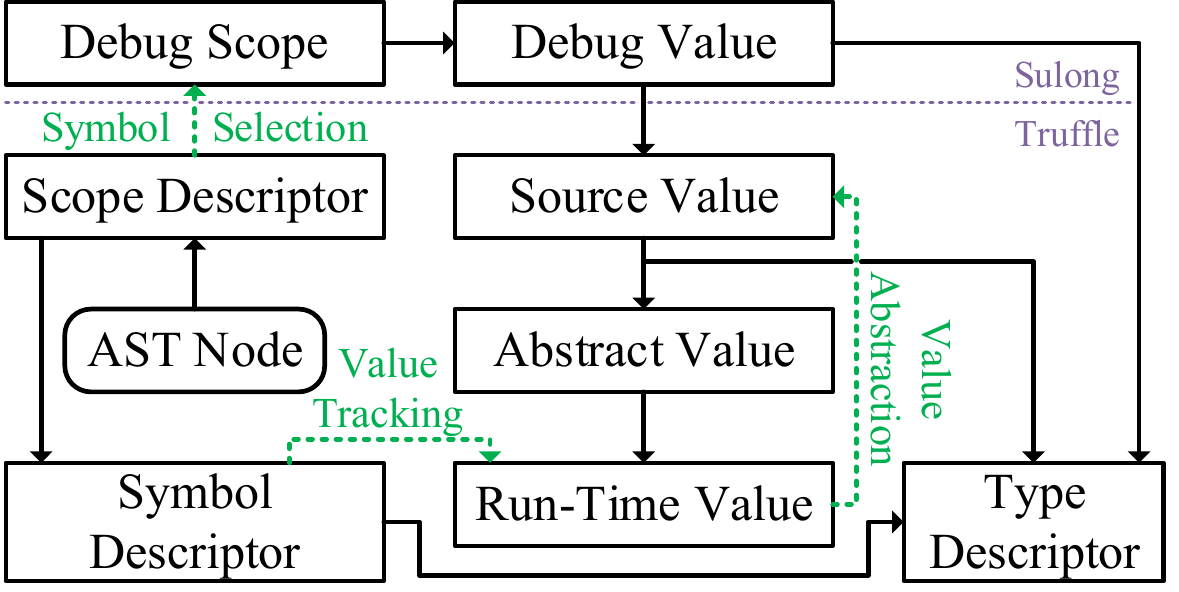}
	\caption{Abstracting from LLVM-IR and Sulong's run-time state to provide access to source-level scopes to the debugger framework.}
	\label{fig:symbols}
\end{figure}

The Truffle debugging framework defines \emph{Debug Scopes} and \emph{Debug Values} as the representation of source-level scopes and their members which it passes to debugger front-ends to display to the user.
Figure~\ref{fig:symbols} shows how Sulong uses symbol, scope and type descriptors to derive the source-level program state in that representation.
Debug values wrap values provided by Truffle language implementations and associate them with a meta-object which contains the name of the value's type for the debugger front-end to display.
A debug scope provides a debug value for each member defined in it.
When the debugging framework retrieves the source-level scope hierarchy at a statement node at which the guest-language program is suspended, the interpreter traverses the hierarchy of scope descriptors attached to the node.
For each scope descriptor, it builds a debug scope containing all symbols defined at the node's location in the original program.
The corresponding debug values wrap \emph{Source Values}, Sulong's representation of source-level values which abstract from LLVM-IR values and the interpreter's representation of them.

\subsection{Value Tracking}

As Sulong executes LLVM-IR, named symbols in the original program can switch between various representations and storage locations, e.g.,~constants, SSA values on the stack, global variables or native memory.
Sulong tracks these changes so it can provide correct values for all symbols to the debugger.
At the LLVM-IR level source-level static symbols never change from their representation as a single global variable.
This is statically encoded in debug information, therefore Sulong does not need to track these symbols at run-time.
LLVM inserts calls to the intrinsic functions \jkCodeEntry{dbg.declare} and \jkCodeEntry{dbg.value} into LLVM-IR functions wherever a local variable changes its representation at the LLVM-IR level.
These intrinsic functions do not have an implementation and calls to them are not meant to be compiled like a regular function call.
Their only purpose is to link debug information and regular program code.
By passing debug information as arguments to a call to \jkCodeEntry{dbg.declare} or \jkCodeEntry{dbg.value}, LLVM indicates that this specific part of debug information is valid at run-time only after the call would have been executed.
These calls are themselves part of debug information and receive the symbol descriptor for the symbol which receives a new value and the symbol's new IR-level value as arguments.
Besides LLVM-IR level global variables and dynamic SSA-values, new values can also be constants.
This allows LLVM to track even those source-level symbols that are not explicitly present in the program, e.g.,~an index variable that was removed during loop unrolling.

In Figure~\ref{fig:fact_AST}, the local variable \jkCodeEntry{n} is assigned only in the first basic-block, which is executed only once.
Sulong detects such effectively-final symbols and stores their values either directly or by reference.
The local variable \jkCodeEntry{result}, on the other hand, receives a value at three points in the program.
This forces Sulong to actively track the current value at run-time which can impose a significant impact on execution time at higher optimization levels.
However, at \jkCodeEntry{-O0}, this overhead is minimal as each source-level local variable lives on the stack, where Sulong needs to update its value to correctly execute the program, and where it is only referenced once by \jkCodeEntry{dbg.declare}.
We believe that programmers typically debug programs without optimizations or at low optimization levels, because optimizations can transform the program in ways that restricts the amount of debug information that can be provided.
Per default, TruffleRuby and GraalPython do not compile native extensions with a higher optimization level than \jkCodeEntry{-O1}.

\subsection{Value Abstraction}

Sulong uses various data structures to represent LLVM-IR level values, ranging from Java primitives to custom classes.
Figure~\ref{fig:symbols} collectively refers to them as \emph{Run-time Values}.
Sulong defines the \emph{Abstract Value} interface as a common way to access these values that is similar to accessing a sequence of bits.
Since source-level variables often live on the heap at the LLVM-IR level, we also implemented this interface for native memory.
A \emph{Source Value} interprets abstract values in a manner determined by a type descriptor.
Using these abstractions, Sulong can represent both primitive values and structured values with an arbitrary number of fields.

\subsection{Symbol Selection}

Sulong's scope descriptors contain any symbol defined within the scope's lexical range.
However, the debugging framework expects to receive only those symbols in a scope that are accessible at the point in the program at which it is suspended.
To avoid displaying symbols that are inaccessible or without a defined value at the suspended statement, Sulong excludes these symbols from the debug scopes it passes to the debugging framework.
Most notably, Sulong excludes any dynamic symbol from debug scopes unless the symbol's declaration site lexically precedes the statement at which the program is halted.
Source-level static symbols, which include all symbols defined in a global scope, have a value at any point in a function since it is preserved across function calls.
Sulong always includes them in debug scopes.

\section{Case Study}
\label{sec:casestudy}

\begin{figure}
	\centering
	\includegraphics[width=.9\columnwidth,page=5,trim={6.35mm 6.35mm 6.35mm 6.35mm},clip]{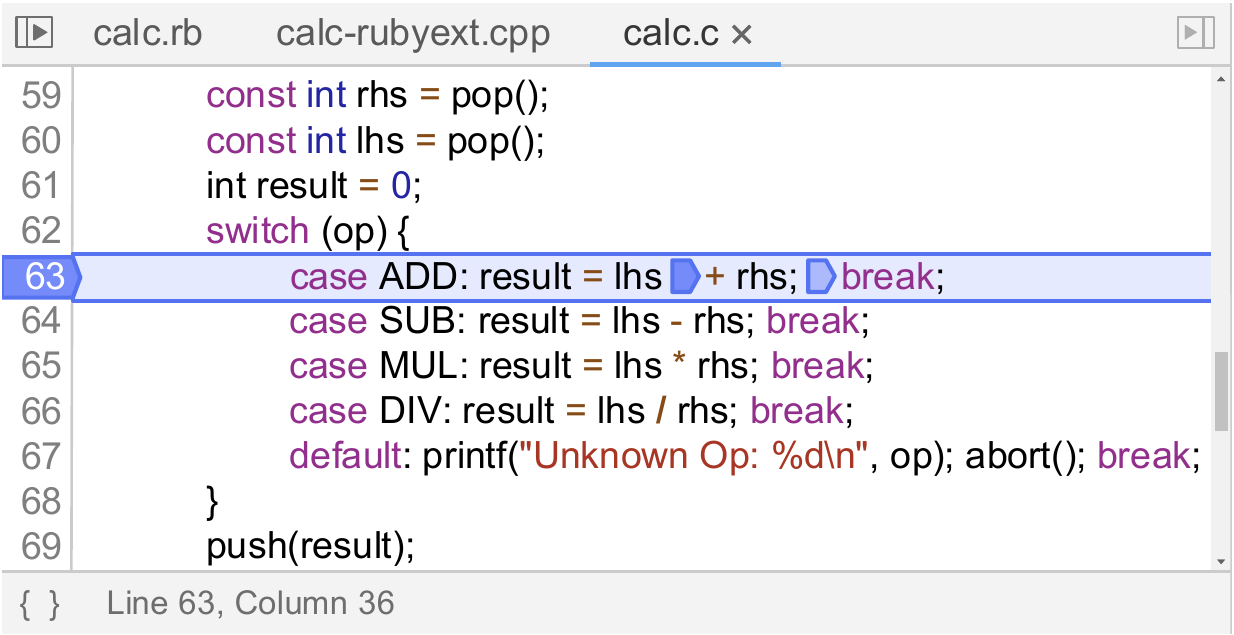}
	\caption{Source View in the Chrome Debugger}
	\label{fig:demo-source}
\end{figure}

\begin{figure}
	\centering
	\includegraphics[width=.6\columnwidth,page=6,trim={6.35mm 6.35mm 6.35mm 6.35mm},clip]{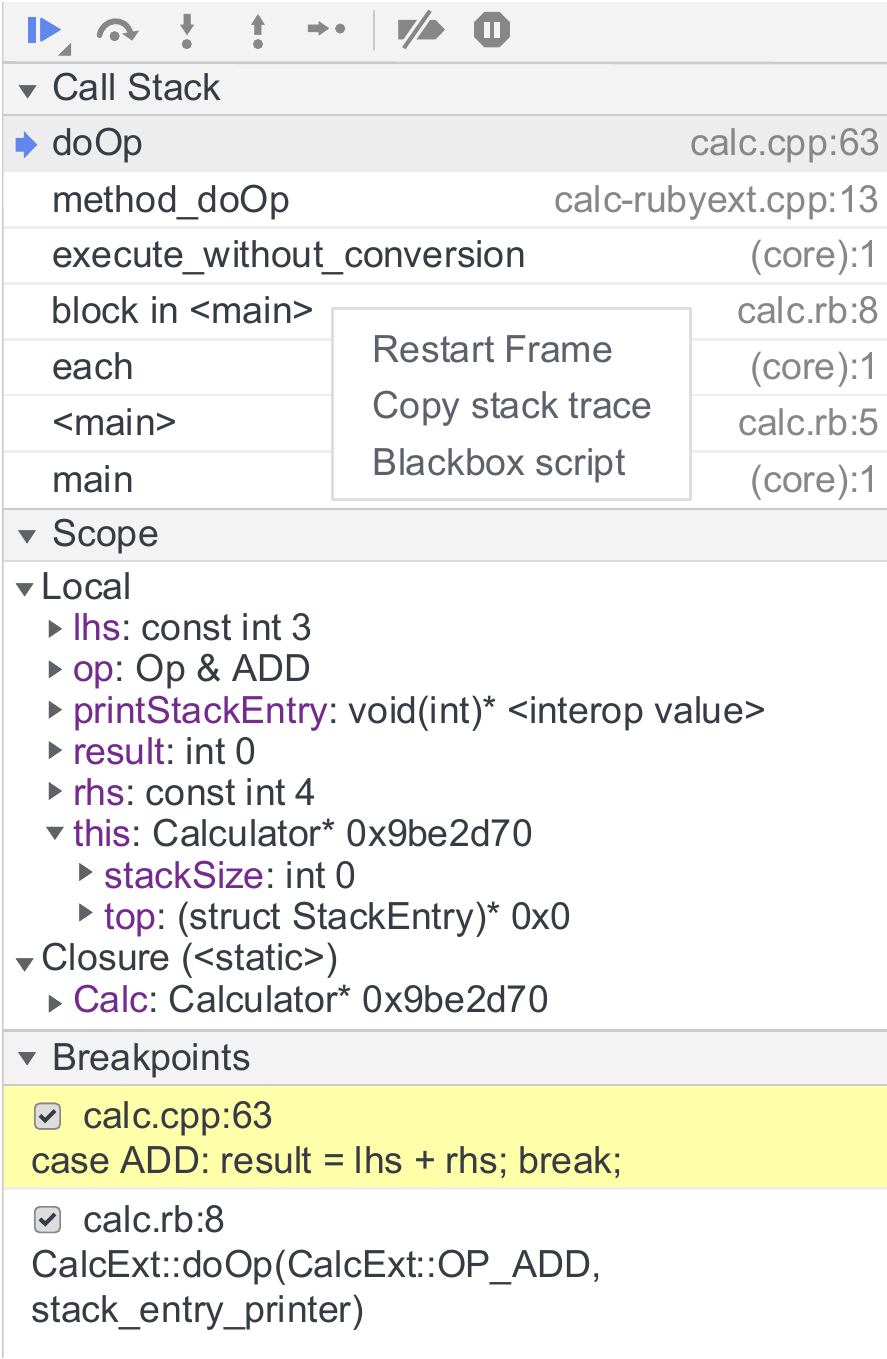}
	\caption{Control View in the Chrome Debugger}
	\label{fig:demo-state}
\end{figure}

\begin{figure}[!h]
	\centering
	\includegraphics[width=.3\columnwidth]{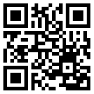}
	\caption{Video Demo at \url{https://youtu.be/iRgL3xycx68}}
	\label{qr:demo1}
\end{figure}

To demonstrate Sulong's debugging support, we implemented a demo program that consists of C++ code that is called by Ruby code.
The complete source code for the demo is available online\footnote{The source code for the demo as well as instructions to debug the code on GraalVM are available at \url{https://github.com/jkreindl/SulongDebugDemo/tree/master/calc}}.
We then used GraalVM to debug this program in the Chrome Debugger.
Figures~\ref{fig:demo-source} and~\ref{fig:demo-state} show the Chrome debugger with the demo program suspended in C++ code that was called from Ruby code.
The referenced code is part of an instance method of a C++ class called \jkCodeEntry{Calculator}.
It pops two numbers off an internal number stack, adds them and pushes the result back onto the stack.
We also recorded a video of this case study where we further detail the application and demonstrate interoperability with Python code (see Figure \ref{qr:demo1}).

\paragraph{Stepping \& Breakpoints}

At the top of Figure~\ref{fig:demo-state}, buttons are shown that enable the user to continue executing the program until the next breakpoint, single-step through the expressions in the source code and into, over and out of function calls.
These stepping modes work also when calling a function defined in another language and passed as parameter to C++ code like the \jkCodeEntry{printStackEntry} variable.
In the figure the program is suspended at the first breakpoint on line 63.
The source view in Figure~\ref{fig:demo-source} shows two breakpoints on line 63 that can be enabled and disabled independently from each other.
Such column-level breakpoints enable users to specify the locations at which they wish the program to be suspended precisely.
All breakpoints that are currently set in the running application can also be seen at the bottom of Figure~\ref{fig:demo-state}.
The presence of a breakpoint on a statement contained in Ruby code further illustrates that in Truffle-based debuggers, both source-level stepping and breakpoints work across language boundaries.

\paragraph{Scopes \& Symbols}

Figure~\ref{fig:demo-state} shows the scope view with two scopes below the call-stack.
The \textit{Local} scope contains all symbols defined within the function denoted by the selected entry on the call-stack.
The full names of their types are provided by the according type descriptors in Sulong.
As can be seen for \jkCodeEntry{lhs} and \jkCodeEntry{rhs}, they also include modifiers such as \jkCodeEntry{const}.
Sulong uses \jkCodeEntry{<static>} as name for the topmost entry in the source-level scope hierarchy, which is the scope formed by a compilation unit.

The local variable \jkCodeEntry{Op} is a C++ reference to an \textit{enum} value.
At the LLVM-IR level it resides in native memory and is described by a pointer to an integer value.
In terms of the abstractions we introduced in Section~\ref{sec:valueinspection}, the corresponding abstract value wraps the native memory denoted by this pointer.
Based on the symbol's type descriptor, the source value then treats this abstract value as an integer and presents it as the label \jkCodeEntry{ADD} to the debug value, which the debugger now displays.
Similarly, \jkCodeEntry{this} is the pointer to the object whose method is currently being called.
This object is a structured value that also resides in native memory.
The corresponding source value uses the pointer's representation as a hexadecimal number as the value to display to the user, but also provides the members defined by the type descriptor to the debugger.
The value of \jkCodeEntry{printStackEntry}, on the other hand, is a complex object defined in Ruby code.
The type descriptor defines it as a pointer to a function, but since it is not actually a pointer, the source value displays it as \jkCodeEntry{<interop value>}.
\jkCodeEntry{lhs}, \jkCodeEntry{rhs} and \jkCodeEntry{result} are signed, 32-bit integer values.

\paragraph{Call-Stack}

In the call-stack, which is shown in the upper half of Figure~\ref{fig:demo-state}, we can see entries for functions implemented in two programming languages, Ruby and C++.
The \textit{.cpp} extension of the corresponding filenames shows that \jkCodeEntry{method\_doOp} and \jkCodeEntry{doOp} are part of C++ code.
The line-numbers shown beside the filenames reference the position in the source code at which the program is currently halted.
In the \jkCodeEntry{doOp} function this corresponds to line 63 which is also highlighted in the code view shown in Figure~\ref{fig:demo-source}, while in \jkCodeEntry{method\_doOp} the number refers to the line in which the function called \jkCodeEntry{doOp}.
The remaining entries on the call-stack reference locations in Ruby code.
Users can select an entry in the call-stack to view the corresponding source code in the code view and the source-level scope hierarchy at the denoted location in the scope view.
The opened context menu in the call-stack also provides the option to \textit{Restart frame}, that is dropping the current state of the function and executing it again from the first instruction.

\section{Conclusion}
\label{sec:conclusion}

In this paper, we have demonstrated our implementation of source-level debugging support in Sulong, an interpreter for LLVM-IR based on the Truffle framework.
This support enables users to debug dynamic programming languages and the native extensions they use in the same debugger front-end.
It is actively being used by TruffleRuby and GraalPython developers.
In contrast to other Truffle-based programming language interpreters, Sulong executes LLVM-IR compiled from various programming languages and uses debug information contained in LLVM-IR to make the source-level program state accessible to a debugger front-end.
We have validated our debugger on applications that consist of Ruby code with C++ extensions and demonstrated interoperability between these languages.
We also recorded a video for the case study we presented in this paper.

\bibliographystyle{ACM-Reference-Format}

\end{document}